# The Role of Thermal Conduction in Tearing Mode Theory

**J W Connor[1,2,3], C J Ham[1], R J Hastie[1,2,3], Y Q Liu[1]**

[1] CCFE, Culham Science Centre, Abingdon, Oxon, OX14 3DB, UK

[2] Imperial College of Science, Technology and Medicine, London SW7 2BZ, UK

[3] Rudolf Peierls Centre for Theoretical Physics, 1 Keble Road, Oxford, OX1 3NP, UK

**Abstract:** The role of anisotropic thermal diffusivity on tearing mode stability is analysed in general toroidal geometry. A dispersion relation linking the growth rate to the tearing mode stability parameter, $\Delta'$, is derived. By using a resistive MHD code, modified to include such thermal transport, to calculate tearing mode growth rates, the dispersion relation is employed to determine $\Delta'$ in situations with finite plasma pressure that are stabilised by favourable average curvature in a simple resistive MHD model. We also demonstrate that the same code can also be used to calculate the basis-functions [C J Ham, et al, Plasma Phys. Control. Fusion **54** (2012) 105014] needed to construct $\Delta'$.

## Introduction

The stability of tearing modes in resistive MHD is very sensitive to the presence of a pressure gradient at the resonant surface [1]. Thus, in toroidal geometry with favourable average curvature (i.e. when the resistive interchange stability parameter, $D_R$, is negative, as can occur in a tokamak) there is a strong stabilising effect at high Lundquist number, S. This is commonly known as the 'Glasser effect' [2] and leads to a critical value, $\Delta'_{crit}$, of the tearing mode stability parameter, $\Delta'$, for instability. This stabilisation is due to the pressure perturbation associated with sound wave propagation. One might therefore expect thermal transport effects in the vicinity of the resonant surface to play an important role in determining the pressure perturbation. Indeed their role has been investigated by Lutjens et al [3, 4]. They found that a length-scale, $w_D = 2\sqrt{2}\left(\chi_\perp/\chi_\parallel\right)^{1/4}\left(Rr_s/ns\right)^{1/2}$, distinct from the resistive layer width, $L_R$, was introduced. Here $\chi_\perp$ and $\chi_{||}$ are, respectively, the thermal diffusivities perpendicular and parallel to the magnetic field, the shear parameter, $s = r\left(dq/dr\right)/q$, n is the toroidal mode number of the tearing mode, $r_s$ the resonant minor radius and R the tokamak major radius. In the situation $r_s >> w_D >> L_R$ Lutjens et al found Glasser stabilisation was essentially replaced by an off-set to the tearing mode stability parameter, $\Delta'$: $\Delta' \to \Delta' - \sqrt{2}\pi^{3/2} D_R/w_D$. The stabilisation described by this off-set is much less than that from the Glasser effect at high S. This suggests that resistive MHD codes, such as MARS-F [5], investigating tearing mode stability should incorporate thermal transport with realistic values for $\chi_\perp$ and $\chi_{||}$, as has been implemented in the XTOR code [6].

An additional interest is the use of such codes to extract a value for $\Delta'$ in toroidal geometry [7, 8]. This is of interest because the physics of the resonant layer model for hot tokamaks will require a kinetic theory description rather than simple resistive MHD. Thus one needs to combine such a resonant layer description with matching to the value of $\Delta'$ obtained from the resistive MHD model. One approach is to use a known analytic dispersion relation relating the growth rate of the tearing mode obtained from the resistive MHD code to $\Delta'$ in order to determine the latter. Of course this method is limited if Glasser stabilisation renders the mode stable [7]. Another approach [8] is to construct basis functions from a resistive MHD code. However, the physics inherent in the Glasser effect shields the resonant surface, preventing construction of a large solution in the sense of Newcomb [9], and negating this approach. A third method is to artificially flatten the equilibrium pressure gradient at the resonant surface with a localised axisymmetric perturbation to it, and develop an analytic relationship between the values of $\Delta'$, with and without this perturbation, the former value now no longer susceptible to the Glasser stabilisation [10]. Since the shielding is related to the effect of the plasma response in the presence of an equilibrium pressure gradient, this last approach suggests a role for thermal transport in modifying this in the vicinity of the resonant surface.

To make full use of a toroidal resistive MHD code with thermal transport included, it is necessary to establish the precise relationship between the growth rate and $\Delta'$ in completely general geometry, as in Ref. 2. The treatments in Refs. 3, 4 and 6 use a somewhat simplified physical model and geometry. In this paper we extend them to general toroidal geometry, using the approach in Ref. 2, modified to include thermal conduction. In the limit that thermal conduction dominates the equation of state, the new characteristic scale, $w_D$, in addition to the basic resistive layer width, $L_R$, enters the theory. Nevertheless it continues to be possible to follow a modified version of the methodology of Ref. 2. An analytic result can be obtained in two geometrical situations. One relies on the favourable average curvature, $D_R = E + F + H^2$ being small, but all other relevant geometrical quantities (i.e. E, F, H as defined in Ref. 1 being arbitrary. The other, simpler case, is to assume H = 0, which resembles the result in [7]. If one wished to relax the condition on $D_R$ when $H \neq 0$, a numerical treatment, analogous to that in Ref. 11, would be necessary. The replacement of Glasser stabilisation by the off-set in $\Delta'$ also has the consequence that the shielding of the resonant surface that prevents the application of the basis function approach also disappears.

In Section 2 we develop the general equations describing the resonant layer including resistivity and anisotropic thermal transport, generalising the approach of Greene and Johnson [12]. The calculations behind the results in Sections 2 are quite lengthy, but closely follow those laid out in Ref. 12, respectively. Rather than repeat these calculations at length we merely describe the essential points and emphasize the differences arising from the inclusion of thermal transport coefficients. Some details are presented in Appendix A. In Section 3 we generate the analytic dispersion equation including finite values of H, following the methodology in Ref. 2, emphasizing the differences arising from the inclusion of thermal transport. The simpler special case with H = 0 is presented in Appendix B. The implications for applying the basis function approach for determining $\Delta'$ are discussed in Section 4, while the application of these various ideas in a fully toroidal resistive MHD code like MARS-F [6] appears in Section 5. Finally we draw conclusions in Section 6.

## 2. The Resonant Layer Equations

We adopt non-orthogonal Hamada co-ordinates, V, $\theta$ and $\zeta$, where V is the volume within a flux surface and $\theta$ and $\zeta$ are angle-like variables increasing by unity after one turn about the torus the short way and the long way round; the Jacobean is unity. The equilibrium magnetic field is written as

$$\mathbf{B} = \nabla V \times (\psi' \nabla\theta - \chi' \nabla\zeta) \tag{1}$$

where $\psi(V)$ and $\chi(V)$ are the toroidal and poloidal fluxes, prime denoting a derivative with respect to V and the scalar pressure is $P(V)$. Similarly the current can be expressed as

$$\mathbf{J} = \nabla V \times (I' \nabla\theta - J' \nabla\zeta) \tag{2}$$

The plasma model adopted consists of the single fluid momentum equation, Ohm's law and an equation of state including anisotropic thermal conduction.

$$\frac{\partial p}{\partial t} + \mathbf{v}.\nabla p + \gamma p \nabla.\mathbf{v} + \mathbf{B}.\nabla \frac{\chi_\parallel}{B^2}(\mathbf{B}.\nabla p) + \chi_\perp \nabla_\perp^2 p = 0 \tag{3}$$

where p is the pressure, $\mathbf{v}$ is the plasma velocity and $\gamma$ is the ratio of specific heats. These equations are linearized, with perturbations described in terms of the plasma displacement, $\xi$, the perturbed magnetic field, **b,** and the perturbed pressure, $\delta p$. The linearized equations are:

$$\rho q^2 \xi = (\nabla \times \mathbf{b}) \times \mathbf{B} + \mathbf{J} \times \mathbf{b} - \nabla \delta p \tag{4}$$

$$\mathbf{b} + \frac{\eta}{q} \nabla \times (\nabla \times \mathbf{b}) = \nabla \times (\xi \times \mathbf{B}) \tag{5}$$

$$q(\delta p + \gamma P \nabla.\xi + \xi.\nabla P) - \nabla.(\chi_\perp \nabla \delta p) - \mathbf{B}.\left[\frac{\chi_\parallel}{B^2}(\mathbf{B}.\nabla \delta p + \mathbf{b}.\nabla P)\right] = 0 \tag{6}$$

where the time variation of all variables has been written as exp (qt).

We represent the displacement and perturbed magnetic field in the form

$$\xi = \xi \frac{\nabla V}{|\nabla V|^2} + \mu \frac{\mathbf{B} \times \nabla V}{B^2} + \nu \frac{\mathbf{B}}{B^2} \tag{7}$$

$$\mathbf{b} = b \frac{\nabla V}{|\nabla V|^2} + v \frac{\mathbf{B} \times \nabla V}{B^2} + \tau \frac{\mathbf{B}}{B^2} \tag{8}$$

The equations are expanded in a narrow layer in the vicinity of a resonant surface, $V = V_s$ where $\iota/2\pi = \chi'/\psi' = n/m$ and the shear parameter is $\Lambda = \psi'\chi'' - \chi'\psi'' = \psi'^2 \iota'/2\pi$ . A new co-ordinate is introduced: $u = \psi_0'\theta - \chi_0'\zeta$, where the subscript 0 corresponds to the value at the resonant surface, so that $\mathbf{B}.\nabla u = \Lambda x/\phi$ with $x = V - V_s$ and $\phi = \psi'/m = \chi'/n$. We make an ansatz for the variation with u: i. e. $\sim \exp(i\alpha u)$, $\alpha >> 1$, but this condition on $\alpha$ can be relaxed in an axisymmetric torus where the perturbations vary as $\exp(2\pi i n u/\chi')$. An ordering scheme based on a small resistivity is introduced:

$$\eta \sim \chi_\perp \sim \varepsilon^8,\ \chi_\parallel \sim \varepsilon^{-2},\ q \sim \varepsilon^2,\ x \sim \varepsilon^3,\ \frac{\partial}{\partial V} \sim \varepsilon^{-3},\ \frac{\partial}{\partial u} \sim \varepsilon^{-1}, \frac{\partial}{\partial \zeta} \sim 1 \qquad (9)$$

and the dependent variables are expanded as

$$\xi = \varepsilon^2\xi^{(2)} + ....,\quad \mu = \mu^{(0)} + ....,\quad \nu = \nu^{(0)} + ....,\quad b = \varepsilon^4 b^{(4)} + ....,\quad v = v^{(0)} + ....,\quad \tau = \varepsilon^2\tau^{(2)} + .... \qquad (10)$$

However we differ from Ref. 12 in the ordering of $\nabla.\xi$. The induction equation (5) requires the ordering $\nabla.(B^2\xi_\perp) \sim \varepsilon^2$ and in the absence of thermal conduction in the equation of state this was not inconsistent with the ordering $\nabla.\xi = \varepsilon^2(\nabla.\xi)^{(2)}$ used in Ref. 12. Now, however, there is an $0(1)$ contribution to $\nabla.\xi$, producing a modification to the $0(\varepsilon^2)$ equation of state that competes with parallel thermal conduction. Physically it allows a competition between sound waves and parallel thermal conduction in determining pressure balance over a connection length.

These expansions are introduced into eqns. (4-6), recalling also that $\nabla.\mathbf{b} = 0$, and the hierarchy of equations solved order by order, as in Appendix B of Ref. 12, averaging over the angle $\zeta$, corresponding to a flux-surface-average, to annihilate unwanted information. In Appendix A we draw attention to the modifications to Ref. 12 that arise from the introduction of thermal transport and the revised ordering of $\nabla.\xi$. Eventually one arrives at a set of equations for certain flux-surface- averaged quantities:

$$\Psi = \langle \mathbf{b}.\nabla V\rangle / i\alpha\Lambda X_0,\ \ \Xi = \langle \xi.\nabla V\rangle,\ \ T = \langle \mathbf{b}.\mathbf{B}\rangle / P' \qquad (11)$$

namely:

$$\Psi_{XX} - HT_X = Q(\Psi - X\Xi) \qquad (12)$$

$$Q^2\Xi_{XX} - QX^2\Xi + (E+F)T + QX\Psi + H\Psi_X = 0 \qquad (13)$$

$$Q(1+\hat{\chi}_\perp M_\infty/M_0)\frac{(1+\hat{\chi}_\parallel Q)}{(1+\hat{\chi}_\parallel QM_\infty/M_0)}T_{XX} - X(1+\hat{\chi}_\parallel Q)(XT - \Psi)$$
$$= Q^2\left[(G+KF)T - (G-KE)\Xi + KH\Psi_X\right] \qquad (14)$$

where

$$E = \frac{\left\langle B^2/|\nabla V|^2 \right\rangle}{\Lambda^2}\left[ J'\psi'' - I'\chi'' + \Lambda \frac{\left\langle \sigma B^2 \right\rangle}{\left\langle B^2 \right\rangle} \right], \quad \sigma = \frac{\mathbf{J.B}}{B^2},$$

$$F = \frac{\left\langle B^2/|\nabla V|^2 \right\rangle}{\Lambda^2}\left[ \left\langle \sigma^2 B^2/|\nabla V|^2 \right\rangle - \frac{\left\langle \sigma B^2/|\nabla V|^2 \right\rangle^2}{\left\langle B^2/|\nabla V|^2 \right\rangle} + P'^2 \left\langle \frac{1}{B^2} \right\rangle \right]$$

$$G = \frac{\left\langle B^2 \right\rangle}{M\gamma P}, \quad H = \frac{\left\langle B^2/|\nabla V|^2 \right\rangle}{\Lambda^2}\left[ \frac{\left\langle \sigma B^2/|\nabla V|^2 \right\rangle}{\left\langle B^2/|\nabla V|^2 \right\rangle} - \frac{\left\langle \sigma B^2 \right\rangle}{\left\langle B^2 \right\rangle} \right], \quad K = \frac{\Lambda^2}{MP'^2}\frac{\left\langle B^2 \right\rangle}{\left\langle B^2/|\nabla V|^2 \right\rangle}. \tag{15}$$

and

$$M_0 = M\left(\hat{\chi}_\parallel = 0\right), \quad M_\infty = M\left(\hat{\chi}_\parallel \to \infty\right), \quad \text{where}$$

$$M = \left\langle B^2/|\nabla V|^2 \right\rangle \left[ \left\langle |\nabla V|^2/B^2 \right\rangle + \frac{1}{\left(1+\hat{\chi}_\parallel Q\right)P'^2}\left( \left\langle \sigma^2 B^2 \right\rangle - \frac{\left\langle \sigma B^2 \right\rangle^2}{\left\langle B^2 \right\rangle} \right) \right],$$

$$\hat{\chi}_\parallel = \chi_\parallel \frac{\rho Q_0}{\gamma P}, \quad \text{while} \quad \hat{\chi}_\perp = \chi_\perp \frac{\left\langle B^2/|\nabla V|^2 \right\rangle \left\langle |\nabla V|^2 \right\rangle}{\eta\gamma P M_\infty},$$

$$\text{and } X = \frac{\left(V - V_s\right)}{X_0}, \quad Q = \frac{q}{Q_0}, \quad \text{with} \quad X_0 = \left( \frac{\rho\eta^2 M \left\langle B^2 \right\rangle^2}{\alpha^2 \Lambda^2 \left\langle B^2/|\nabla V|^2 \right\rangle^2} \right)^{1/6} \text{ and } Q_0 = \left( \frac{\eta\alpha^2\Lambda^2 \left\langle B^2 \right\rangle}{\rho M \left\langle B^2/|\nabla V|^2 \right\rangle} \right)^{1/3}, \tag{16}$$

The angle brackets are defined by $\left\langle A \right\rangle = \left(1/2\pi\right)\oint d\theta\, A$. It is also convenient to define the Mercier (ideal MHD) stability index, $D_I$ [13], and 'average curvature', $D_R$ :

$$D_I = E + F + H - 1/4$$

$$D_R = E + F + H^2 \tag{17}$$

We emphasize that the appearance of the ratio $M_\infty / M_0$ in eqn. (14) is a convenient way of representing the exact variation of $M$ with $\hat{\chi}_\parallel$, not an approximate interpolation formula. In the limit $\hat{\chi}_\perp$, $\hat{\chi}_\parallel \to 0$ we recover the equations of Ref. 2. It is interesting to note that the toroidal enhancement of inertia, represented by the quantity $M_0$ , i. e. M when $\hat{\chi}_\parallel = 0$, disappears when $\hat{\chi}_\parallel >> 1$.

In the transport dominated limit: $\hat{\chi}_\perp, \hat{\chi}_\parallel >> 1$, eqn. (14) reduces to the simpler form:

$$\hat{\chi}_\perp T_{XX} = \hat{\chi}_\parallel X(XT - \Psi), \tag{18}$$

or, introducing a new scale-length, $y = (\hat{\chi}_\parallel / \hat{\chi}_\perp)^{1/4} X$,

$$T_{YY} - y^2 T = -\left(\frac{\hat{\chi}_\parallel}{\hat{\chi}_\perp}\right)^{1/4} \Psi y \tag{19}$$

We shall consider this situation in the additional limit $\hat{\chi}_\parallel / \hat{\chi}_\perp << 1$ so that the scale-length associated with eqn. (17) is much longer than that appearing in eqns. (12) and (13). Although $\chi_\parallel >> \chi_\perp$ in physical reality, since

$$\frac{\hat{\chi}_\parallel}{\hat{\chi}_\perp} = X_0^4 \frac{\chi_\parallel}{\chi_\perp} \frac{\alpha^2 \Lambda^2}{\langle B^2 \rangle \langle |\nabla V|^2 \rangle}, \tag{20}$$

the smallness of $X_0 / V_s$ makes this a reasonable assumption.

## 3. The Dispersion Relation for H Finite

The dispersion relation for the tearing mode results from matching the solution of the inner resistive region equations to the external ideal MHD solutions. It involves the tearing mode stability parameter $\Delta'$ that represents the jump in the ratio of small and large solutions [9] across the resonant surface. These solutions behave as

$$\Psi \sim X^p, \ \Xi = T = \Psi / X; \quad p = 1/2 \pm (-D_I)^{1/2} \tag{21}$$

As in Ref. 2 we introduce the ordering: $Q \sim \delta^{2/3}, X \sim \delta^{1/6}$ , implying $\Psi \sim 1, \ \Xi \sim T \sim \delta^{-1/6}$, together with the condition on the equilibrium, $D_R \sim \delta$, while $E \sim F \sim H \sim 1$. In leading order eqns. (12) and (13) are

$$\Psi_{XX}^{(0)} - H T_X^{(0)} = 0 \tag{22}$$

$$(E + F) T_{XX}^{(0)} + H \Psi_{XX}^{(0)} = 0 \tag{23}$$

so that in the present ordering they are no longer independent. To obtain an independent equation we differentiate eqn. (13) and use eqn. (12) to eliminate $\Psi_{XX}^{(0)}$, resulting in an equation for $\Xi^{(0)}$:

$$\left(Q\Xi_{XX}^{(0)} - X^2\Xi^{(0)}\right)_X - HX\Xi^{(0)} + \left(X\Psi^{(0)}\right)_X + H\Psi^{(0)} + \left(D_R/Q\right)T_X^{(0)} = 0 \qquad (24)$$

First we must discuss the nature of the solutions of eqns. (12), (18) and (24) at large X for matching to the ideal region. One is given by

$$\Psi \sim X^H, \;\; \Xi \sim T \sim \Psi/X \sim X^{H-1}; \qquad (25)$$

however, it is necessary to make $\Xi$ an order larger in $\delta$ to identify the other solution:

$$\Psi \sim X^{1-H}, \; T \sim \Psi/X, \;\; \Xi \sim H(1-2H)X^{-2-H}/Q\,. \qquad (26)$$

This solution for $\Xi$ fails to match to the ideal solutions (21), (since $p = H$ or $1\text{-}H$ when $D_R \sim \delta$), but consideration of an intermediate region enables a smooth matching. Thus, in solving the resonant layer equations we need to identify the components corresponding to the asymptotic forms (25) and (26).

We return to the solution of the resonant layer equations; the lowest order terms in eqns. (12) and (18) yield

$$\Psi_{XX}^{(0)} - HT_X^{(0)} = 0 \qquad (27)$$

$$\hat{\chi}_{\perp}T_{XX}^{(0)} - \hat{\chi}_{\parallel}X^2T^{(0)} + \hat{\chi}_{\parallel}X\Psi^{(0)} = 0 \qquad (28)$$

which reduce to

$$\Psi_{yy}^{(0)} - y^2\Psi^{(0)} + Hy\Psi^{(0)} = 0 \qquad (29)$$

precisely the same equation for $\Psi^{(0)}$ as obtained in Ref. 2, but with the variable $y = \left(\hat{\chi}_{\parallel}/\hat{\chi}_{\perp}\right)^{1/4}X$ replacing $z = X/Q^{1/4}$. Thus

$$y = \lambda z, \;\; \lambda = \left(Q\hat{\chi}_{\parallel}/\hat{\chi}_{\perp}\right)^{1/4} \qquad (30)$$

We can therefore follow the same solution procedure to obtain

$$\Psi^{(0)}(z) = \Psi_1(z) + \Psi_2(z) \qquad (31)$$

with

$$\begin{aligned}\Psi_1(z) &= \lambda^{-1}\exp(2\nu\pi i)\int_{C_1} dk\exp(ikz)\left(s/\lambda^2\right)^{-\nu}K_\nu(s\exp(-2\pi i)/\lambda^2);\\ \Psi_1(z) &= \lambda^{-1}\exp(-2\nu\pi i)\int_{C_2} dk\exp(iky)\left(s/\lambda^2\right)^{-\nu}K_\nu(s\exp(2\pi i)/\lambda^2),\end{aligned} \qquad (32)$$

where $C_1$ is a contour in the k-plane from $\infty \exp(3i\pi/4)$ to $\infty \exp(i\pi/4)$, entirely in the upper half plane, while $C_2$ has the same limits but passes below the origin, $s = k^2/2$ and $\nu = (1+H)/4$. The solutions (32) follow from forming the Fourier transform in a variable $\kappa = k/\lambda$ with respect to y and then re-expressing it in terms of z. The asymptotic form of $\Psi^{(0)}$ is

$$\Psi^{(0)}(z) \sim \Psi_2(z) \sim 2^{3\nu}\pi[\Gamma(\nu)/\Gamma(4\nu)](\lambda z)^{4\nu-1}[1+0((\lambda z)^{-4})] \tag{33}$$

To obtain the other power-like behaviour at large z (i. e. large X), we must consider eqn. (24) which becomes

$$\left(\Xi^{(0)}_{zz} - z^2\Xi^{(0)}\right)_z - Hz\Xi^{(0)} + Q^{-1/4}\left[\left(z\Psi^{(0)}\right)_z + H\Psi^{(0)} + \left(D_R/HQ^{3/2}\right)\Psi^{(0)}_{zz}\right] = 0. \tag{34}$$

For the appropriate odd-solution for $\Xi^{(0)}$ we require the inhomogeneous solution of eqn. (34). This is achieved by introducing the Fourier transform and solving the resulting second order equation in k by the method of variation of parameters. However we can readily identify part of the solution at large z from the dominant terms in eqn. (34):

$$\Xi^{(0)}_1 \sim \Psi^{(0)}(z)/zQ^{1/4}, \tag{35}$$

However this fails to generate the solution (26). The part of the solution for $\Xi^{(0)}$ that does produce the necessary powers in z is given by

$$\Xi^{(0)} = \int_{C_2} dk \exp(ikz)s^{\nu}$$
$$\times\left\{K_\nu(s)\int_{\exp(-2\pi i)}^{0} dt\,t^{-\nu}K_\nu\left(t\exp(2\pi i)\right)R\left(t\exp(2\pi i)\right) - \exp\left(-4\pi\nu i\right)K_\nu(s\exp(2\pi i))\int_0^\infty dt\,t^{-\nu}K_\nu\left(t\right)R\left(t\right)\right\} \tag{36}$$

where

$$R(t) = \frac{\exp(2\nu\pi i)t^{-\nu+1/2}}{\left[2\sqrt{2}\pi Q^{1/4}\cos\left(\nu\pi\right)\lambda^{(2\nu-1)}\right]}\left[\frac{1}{\lambda^2}K'_\nu\left(\frac{t}{\lambda^2}\right) + \frac{(1/2-\nu)}{t}K_\nu\left(\frac{t}{\lambda^2}\right) + \left(\frac{D_R}{HQ^{3/2}}\right)K_\nu\left(\frac{t}{\lambda^2}\right)\right] \tag{37}$$

This follows from eqn. (68) of Ref. 2 after one allows for the different scale-length that controls $\Psi^{(0)}$; the remaining terms in that equation merely contribute to solution (35).

To obtain the asymptotic form for $\Xi^{(0)}$ we first evaluate the definite integrals which both take the form $\int_0^\infty dt\,t^{-\nu}K_\nu(t)R(t)$. We express $K'_\nu(t)$ in terms of $K_\nu(t)$ and $K_{\nu+1}(t)$ and use the result **7.14.2** (36) of Ref. 14:

$$2^{\rho+2}\Gamma(1-\rho)\int_0^\infty dt t^{-\rho}K_\mu(\alpha t)K_\nu(\beta t) = \alpha^{\rho-\nu-1}\beta^\nu$$
$$\times\ {}_2F_1\left((1+\nu+\mu-\rho)/2,(1+\nu-\mu-\rho)/2;1-\rho;1-\beta^2/\alpha^2\right)$$
$$\times\ \Gamma\left((1+\nu+\mu-\rho)/2\right)\Gamma\left((1+\nu-\mu-\rho)/2\right)\Gamma\left((1-\nu+\mu-\rho)/2\right)\Gamma\left((1-\nu-\mu-\rho)/2\right) \tag{38}$$

where ${}_2F_1$ is the hyper-geometric function [15], setting $t=\lambda^2 u$, $\alpha=1$ and $\beta=\lambda^2$, with appropriate choices of $\mu$ and $\nu$. In taking the limit $\lambda<<1$, when the final argument of the hyper-geometric function becomes $1\text{-}\beta^2/\alpha^2=\lambda^4$, one uses the transformation formula **15.3.6** formula of Ref. 15 which generates two terms, corresponding to different powers of $\lambda$. Alternatively one can expand the term in $K_\nu(\beta u^2)$, retaining the two lowest powers in $\beta u^2$, and then using the result **7.7.4** (27) of Ref. 14, thus again generating two powers of $\lambda$.) This differs from the situation in Ref. 2 when $\lambda=1$, the argument $1\text{-}\beta^2/\alpha^2=0$ and only one term is generated.

For large z we can use the small argument expansion of $K_\nu(s)$ and the resulting integral over k can be recognised as the Hankel representation of a Gamma function [16]. Consequently the asymptotic form for $\Xi^{(0)}$ can be written as

$$\Xi^{(0)} \sim \Psi^{(0)}(z)/zQ^{1/4} + C_\nu z^{-H-2}, \tag{39}$$

where

$$C_\nu = -\frac{\pi}{2^{5\nu+1}}\frac{\cos(2\pi\nu)\Gamma(1/4-\nu)\Gamma(-\nu)\Gamma(1/4)}{\Gamma(\nu+1)\Gamma(-4\nu)}$$
$$\times\left\{\nu-\frac{D_R}{HQ^{3/2}}\left[\frac{\Gamma(3/4-\nu)\Gamma(3/4)}{\Gamma(1/4-\nu)\Gamma(1/4)}\lambda^2+\frac{\Gamma(3/4-2\nu)\Gamma(3/4-\nu)\Gamma(\nu)}{\Gamma(1/4-\nu)\Gamma(1/4)\Gamma(-\nu)}\lambda^{2-4\nu}\right]\right\} \tag{40}$$

We note the appearance of two distinct powers of $\lambda=\left(Q\hat{\chi}_\parallel/\hat{\chi}_\perp\right)^{1/4}$.

To obtain the complete solution for $\Psi(z)$ we require $\Psi^{(1)}(z)$. In first order, eqns. (12) and (14) yield

$$\Psi^{(1)}_{XX} - HT^{(1)}_X = Q\left(\Psi^{(0)} - X\Xi^{(0)}\right) \tag{41}$$

$$\hat{\chi}_\perp T^{(1)}_{XX} = \hat{\chi}_\parallel X\left(XT^{(1)}_X - \Psi^{(1)}\right) \tag{42}$$

Solving these for large X we obtain, in terms of z,

$$\Psi^{(1)}(z) \sim \frac{Q^{7/4}}{H(1-2H)}C_\nu z^{1-H} \tag{43}$$

The external ideal MHD solution has the form

$$\Psi_{L,R} = A_{L,R}|Y|^{H} + B_{L,R}|Y|^{1-H}; \quad Y = (V - V_s)/V_s = \left(X_0 Q^{1/4}/V_s\right) z \tag{44}$$

and if we write the inner resonant layer solution as

$$\Psi_I = A_I|Y|^{H} + \left(B_I + \tilde{B}_I \operatorname{sgn}(Y)\right)|Y|^{1-H}, \tag{45}$$

where $A_I$ and $B_I$ are obtained from eqns. (33) and (43) and $\tilde{B}_I$ is an arbitrary coefficient arising from an even solution, $\Xi^{(0)}$, to the homogeneous eqn. (34). The matching condition is

$$A_L = A_R = A_I, \quad B_L = B_I - \tilde{B}_I, \quad B_R = B_I + \tilde{B}_I \tag{46}$$

Defining

$$\Delta = B_R/A_R + B_L/A_L \tag{47}$$

and

$$\Delta(Q) = 2B_I/A_I \tag{48}$$

we finally obtain the dispesion relation

$$\Delta' = \Delta(Q) \tag{49}$$

with

$$\Delta(Q) = \frac{\pi}{4}\frac{Q^{(5+2H)/4}}{((1-2H))}\left(\frac{2V_s}{X_0}\right)^{1-2H}\left(\frac{\hat{\chi}_\perp}{\hat{\chi}_\parallel}\right)^{H/4}\frac{\cos(\pi H/4)}{\sin((\pi(1+H)/2)}\frac{\Gamma(1/4)\Gamma((3-H)/4)\Gamma(1+H)}{\Gamma((1+H)/4)\Gamma((5+H)/4)\Gamma(1+H/4)}$$
$$\times\left\{1+H+\frac{D_R}{Q}\left(\frac{\hat{\chi}_\parallel}{\hat{\chi}_\perp}\right)^{1/2}\frac{\Gamma(3/4)\Gamma((2-H)4)}{\Gamma(1/4)\Gamma(1-H/4)}-\frac{D_R}{Q^{(5+H)/4}}\left(\frac{\hat{\chi}_\parallel}{\hat{\chi}_\perp}\right)^{(1-H)/4}\frac{\Gamma((1-2H)/4)\Gamma((2-H)/4)\Gamma((5+H)/4)}{\Gamma(1/4)\Gamma(1-H/4)\Gamma((3-H)/4)}\right\}$$

(50)

In the limit H = 0 this reduces to

$$\Delta'(Q) = 2\pi\frac{V_s}{X_0}Q^{5/4}\frac{\Gamma(3/4)}{\Gamma(1/4)}\left[1+\sqrt{\pi}\frac{\Gamma(3/4)}{\Gamma(1/4)}\left(\frac{\hat{\chi}_\parallel}{\hat{\chi}_\perp}\right)^{1/2}\frac{D_R}{Q}\right]-\frac{V_s}{X_0}\frac{\pi^{3/2}}{2}\left(\frac{\hat{\chi}_\parallel}{\hat{\chi}_\perp}\right)^{1/4}D_R\,. \tag{51}$$

This expression can be compared with that of the resistive MHD model [2]

$$\Delta(Q) = 2\pi\frac{V_s}{X_0}Q^{5/4}\frac{\Gamma(3/4)}{\Gamma(1/4)}\left[1-\frac{D_R}{Q^{3/2}}\right]. \tag{52}$$

In eqn. (49) the quantity $\Delta'$ is calculated from the marginal ideal MHD equations that satisfy appropriate boundary conditions at the magnetic axis and plasma surface. However the equations including thermal conduction differ from the standard MHD equations, leading to a different linearized equation for the perturbed pressure, $\delta p$. It might therefore be conjectured that the value of $\Delta'$ has also been altered by the role of strong parallel thermal conduction. We now show that this is not the case by examining solutions of an equation of state dominated by parallel thermal conduction rather than compressibility:

$$\mathbf{B}.\nabla\left[\frac{\chi_{\|}}{B^2}\left(\mathbf{B}.\nabla\delta p + \mathbf{b}.\nabla P\right)\right] = 0 \tag{53}$$

where P(V) is the equilibrium scalar pressure. In the presence of magnetic shear, i.e. for irrational magnetic surfaces, eqn. (53) can be integrated to yield $(\mathbf{B}.\nabla p + \mathbf{b}.\nabla P) = 0$, where we have invoked the solubility condition on this magnetic differential equation [17]. Since $\mathbf{b} = \nabla \times (\boldsymbol{\xi} \times \mathbf{B})$, this can be written $\mathbf{B}.\nabla(p + P'\xi) = 0$. Integrating once again, $p = -P'\xi$, where we set the constant of integration to zero as the perturbations depend on $\theta$ and $\zeta$. This is the same result as standard ideal MHD so that it leads to the same value for $\Delta'$.

## 4. Calculations of $\Delta'$ from the Dispersion Relation using the MARS-F Code

We have calculated $\Delta'$ for a toroidal equilibrium which is stable due to the Glasser effect in the resistive MHD model for $S > 10^9$, by replacing the equation of state in the MARS-F code with the one including transport processes. In the limit that transport dominates one can take the growth rate calculated from the code and exploit the analytic dispersion relation (50) to determine $\Delta'$. The finite $\beta$ equilibrium chosen has $R/a = 10,\ q_0 = 1.44,\ q_a = 2.94,\ \beta = 0.019\%$. It has the properties $D_R = -3.5\times10^{-3},\ H = 7.9\times10^{-4}$ and a single resonant surface at $q = m/n = 2$. Figure 1 compares the growth rate as $\chi_{\|}/\chi_{\perp}$ is increased when only transport effects are retained in the equation of state, with the result without them, showing how transport overcomes Glasser stabilisation at high values of the Lundquist number, S . In Fig. 2 we use the dispersion relation (50) to deduce $\Delta'$ from the growth rate as a function of $\chi_{\|}/\chi_{\perp}$ for a high Lundquist number, $S = 10^9$, in the transport only case, showing the effect of H. The results are insensitive to $\chi_{\|}/\chi_{\perp}$ and the outcomes from the two dispersion relations (50) and (51) almost lie on top of each other since $H = 7.9\times10^{-4} << 1$. Figure 3 shows the convergence of $\Delta'$ with S for $\chi_{\|}/\chi_{\perp} = 10^2$; its value asymptotes to a constant for $S \geq 10^9$. The result obtained using the resistive MHD dispersion relation of Ref. 2 is also shown since complete Glasser stabilisation only operates for $S > 10^9$.

## 5. Basis-functions

In an earlier publication [8] a method for calculating $\Delta'$ in a torus using a set of basis-functions to be obtained from a toroidal resistive MHD code was described. In the simplest case of one resonant surface this involved constructing, as a stationary solution of the resistive MHD code, a small solution with its set of accompanying harmonics emerging from its resonant surface, the resonant harmonic having unit slope and the other harmonics zero slope. Each harmonic of this solution has finite amplitude at the wall. We associate with this solution a second basis-function, the 'response' solution, which is calculated by imposing zero amplitude on all harmonics at the magnetic axis and the values of the original small solution at the wall; the harmonics are allowed to reconnect (i.e. they are continuous) across their resonant surfaces. Subtracting these two solutions yields a solution satisfying the appropriate boundary conditions, from which one can calculate $\Delta'$ as the ratio of the jump in the small solution across the resonant surface to the amplitude of the large solution there. The method readily generalises to multiple resonant surfaces.

Unfortunately the approach is only satisfactory when there is zero pressure gradient at the resonant surface because otherwise, within a resistive MHD model, the stationary response function is shielded from the resonant surface and one cannot identify an amplitude for the large solution there. Including thermal transport effects in the toroidal resistive MHD code can be expected to resolve this problem as the response solution can penetrate to the resonant surface, essentially for the same reason that Glasser stabilisation has been removed.

To see this we consider the stationary limit, $Q \to 0$, of eqns. (12)-(14):

$$D_R \Psi_{XX} - HT_X = 0 \tag{54}$$

$$(E+F)T + H\Psi_X = 0 \tag{55}$$

$$Q(1+\hat{\chi}_\perp)T_{XX} - X^2(1+\hat{\chi}_\parallel Q)(XT-\Psi) = 0 \tag{56}$$

When $\hat{\chi}_\perp,\ \hat{\chi}_\parallel << 1$, eqn. (56) reduces to

$$T = \Psi/X \tag{57}$$

whereas, if $\hat{\chi}_\perp,\ \hat{\chi}_\parallel >> 1$,

$$\hat{\chi}_\perp T_{XX} - X^2 \hat{\chi}_\parallel (XT - \Psi) = 0 \tag{58}$$

If $D_R \neq 0$ it follows from eqns. (54), (55) and (57) that $D_R\left(\Psi_X/X - \Psi/X^2\right) = 0$, implying $\Psi \propto |X|$, i.e. $\Psi(0) = 0$. Thus the resonant surface is shielded from a stationary external perturbation of $\Psi$.

However if transport dominates we only learn from eqns. (54), (55) and (58), that $\Psi_{XX}(0)=0$, allowing a finite value of $\Psi(0)$, i.e. penetration is allowed.

Furthermore eqns. (12) - (14) reduce to those of ideal MHD away from the resonant surface. Thus for small values of the perpendicular thermal diffusivity, it only plays a role near the resonant surface. On the other hand parallel transport is important, dominating eqn. (14). The result is $XT=\Psi$, but this is what is obtained in the ideal MHD model, since the equation of state (14) then leads to $XT=\Xi$, while the induction equation (12) implies $\Xi=\Psi$. Consequently a resistive MHD code (e.g. Ref. 6) including thermal transport can be used to construct appropriate basis functions for calculating $\Delta'$.

The MARS-F code has been used to construct the appropriate toroidal basis-functions and deduce $\Delta'$. Figure 4 presents the basis-functions for several poloidal harmonics for the resistive MHD model and Figure 5 those for the transport dominated model, showing how the m = 2, n = 1 resonant harmonic is screened in the former case, whereas it can penetrate to the resonant surface in the latter. Figure 6 and 7 show the resulting convergence of $\Delta'$, calculated by the basis function method, with $\chi_{\|}/\chi_{\perp}$ and S, respectively, for the same equilibrium as used for Fig. 1. These figures also include the results from Figures 2 and 3 obtained using the dispersion relation method for the resistive MHD and transport models for comparison.

## 6. Discussion and Conclusions

Equations (49) and (50) provide a dispersion relation for tearing modes in general toroidal geometry for a resistive plasma model including anisotropic transport of pressure. Compared to the structure in the absence of thermal conduction, eqn. (52) as obtained in Ref. 2, there are now three terms inside the curly bracket rather than just two. The second term, proportional to $\left(\hat{\chi}_{\|}/\hat{\chi}_{\perp}\right)^{1/2}$ has a destabilising effect when there is favourable average curvature, $D_R<0$; in this case, the third term, proportional to $\left(\hat{\chi}_{\|}/\hat{\chi}_{\perp}\right)^{1/4}$ and thus larger than the second, has the nature of a stabilising off-set to the tearing mode stability parameter $\Delta'$. Of course the opposite conclusions arise when $D_R>0$, as in a Reversed Field Pinch.

The first two terms have the appearance of a modification of the two terms in Ref. 2 which describe the influence of the interchange mode, but now has the feature that these transport processes undercut the Glasser stabilisation that is present in toroidal plasma with favourable average curvature and a finite pressure gradient at the resonant surface. This has the positive aspect that it means one can use this dispersion relation to determine a toroidal $\Delta'$ from tearing mode growth rates produced in a toroidal resistive MHD code incorporating transport of pressure, without the problem that Glasser stabilisation prevents their growth. However there remains the stabilising off-set from the third term. This leads to the tearing mode stability criterion (taking H = 0, for simplicity)

$$\Delta' < \Delta'_{crit} = \frac{\pi^{3/2}}{2}\left(\frac{\chi_{\|}}{\chi_{\perp}}\right)^{1/4} V_s \left(\frac{\alpha^2\Lambda^2}{\langle B^2\rangle\langle|\nabla V|^2\rangle}\right)^{1/4} D_R \tag{59}$$

In a large aspect ratio tokamak we note that, using results in Ref. 18, but recalling that the toroidal enhancement of inertia is supressed when $\hat{\chi}_\parallel >> 1$,

$$\frac{\hat{\chi}_\parallel}{\hat{\chi}_\perp} = \left(\frac{ns}{r_s R}\right)^{1/3} \left(\rho \frac{\eta^2}{B^2}\right)^{2/3} \frac{\chi_\parallel}{\chi_\perp} \tag{60}$$

and

$$\Delta'_{crit} = -\frac{\pi^{3/2}}{2} \left(\frac{\chi_\parallel}{\chi_\perp}\right)^{1/4} \left(\frac{ns}{Rr_s}\right)^{1/2} D_R \tag{61}$$

If one considers the effective value of $\Delta'$, i.e. after subtracting the stabilising off-set $\Delta'_{crit}$, to be small, the dispersion relation admits an unstable interchange-like mode, but driven by $D_R < 0$; however the growth rate is very weak: $q \propto (\chi_\parallel/\chi_\perp)^{1/2}/\tau_R$ where $\tau_R$ is a resistive diffusion time.

Another feature of the new dispersion relation is that both these second and third terms have the property that they do not produce a diverging contribution to $\Delta'$ as $Q \to 0$, unlike the case in Ref. 2. (This is true provided H > 0, but normally this is so: Ref. 18 shows this for a large aspect ratio tokamak.) This means that stationary, external, magnetic perturbations can penetrate to their resonant surfaces - i.e. they are not shielded by the presence of a finite pressure gradient and favourable average curvature. (However, one should note that shielding due to plasma rotation may be present.) Not only would this have implications for the application of resonant magnetic perturbations for ELM control and for the effects of error fields, but it also means the procedure for constructing basis-functions for determining $\Delta'$ is viable, since the large solution exists at the resonant surface, without needing to resort to devices like equilibrium pressure flattening.

The dispersion relation has been obtained for the case of small average curvature, $D_R << 1$, though H can be finite. In Appendix B a similar result is obtained for finite $D_R$, but small values of the equilibrium quantity H. If neither of these situations prevails one would need to produce a numerical solution of the basic equations, in the spirit of Ref. 11, but including transport of pressure. In the extreme limit $(\hat{\chi}_\perp/\hat{\chi}_\parallel)^{1/4} >> X_0$, the dispersion relation reduces to that in the absence of pressure gradients (apart from the effect of H in the pre-factor in eqn. (50).

The analysis presented above is based on $(\hat{\chi}_\parallel/\hat{\chi}_\perp) << 1$, corresponding to the pressure flattening width exceeding the resistive layer width, $L_R$. One can readily extend the analysis to arbitrary values of this parameter using the expression (38) for arbitrary $\lambda$. As shown in Appendix B for the case H = 0 (which can extended to general H straightforwardly), the dispersion relation takes the form

$$\Delta(Q) = 2\pi \frac{V_s}{X_0} Q^{5/4} \frac{\Gamma(3/4)}{\Gamma(1/4)} \left[1 - \frac{1}{2^{5/2}\pi^{3/2}} \frac{\Gamma(1/4)}{\Gamma(3/4)} \frac{D_R}{Q^{3/2}}\right] \quad , \tag{62}$$

Resembling that shown in eqn. (52), with the consequence that a Glasser-like stabilisation is restored, if $(\hat{\chi}_\parallel/\hat{\chi}_\perp) >> 1$ when the pressure flattening width is less than the resistive layer width. It is interesting to compare the values for the critical $\Delta'$ arising from eqns. (59) and (62). On appropriately adapting the result in Ref. 2, eqn. (62) leads to

$$\Delta'_{crit} = 0.17(V_s/X_0)|D_R|^{5/6} \tag{63}$$

This can be lower than the prediction of eqn. (58) if $\chi_\parallel/\chi_\perp$ is sufficiently large, resulting in larger growth rates in this situation, provided the Glasser-like stabilisation is overcome. Furthermore it is also lower than the result Ref. 2:

$$\Delta'_{crit} = 1.54(V_s/X_0)|D_R|^{5/6} \tag{64}$$

so that the growth rate will also be larger than in the absence of transport. This difference results from the somewhat different equations of state in the two cases. In the $(\hat{\chi}_\parallel/\hat{\chi}_\perp) >> 1$ case pressure remains constant on a perturbed magnetic field line, rather than satisfying a simple compressible equation of state.

The analysis above was entirely concerned with linear theory. However it is well known that related thermal conduction effects play a key role in non-linear neo-classical island growth [19], defining a critical island width for growth given by

$$w_D/r_s = 2\sqrt{2}(\chi_\perp/\chi_\parallel)^{1/4}(R/nsr_s)^{1/2}. \tag{65}$$

The use of the dispersion relation (50) to determine $\Delta'$ from the growth rate calculated by the MARS-F code, adapted to include a transport model, was demonstrated in Figures 2 and 3 for a specific toroidal equilibrium.

In Section 5 we used the basis-function method to determine $\Delta'$. We first showed that the transport model enables penetration of the response function at a resonant surface even in the presence of a finite pressure gradient and favourable average curvature, unlike the usual resistive MHD equations. This leads to a finite value for the 'large' solution at the resonant surface and hence allows the calculation of $\Delta'$. The method was demonstrated in Figs 4 and 5 for the same equilibrium used for the dispersion relation method. The results for $\Delta'$ were the same as shown in Figs 6 and. 7. We stress the fact that the basis-function method does not rely on the tearing mode being linearly unstable, in contrast to the dispersion relation method, provided the resonant surface is not screened from the response function.

In summary then, we have investigated the effect of anisotropic thermal transport on the stability of tearing modes, obtaining a dispersion relation relating the growth rate to $\Delta'$ in general toroidal geometry. In the limit that transport effects dominate the equation of state the stabilisation due to the Glasser effect is greatly reduced. This enables one to determine $\Delta'$ by using a resistive MHD code, such as MARS-F, adapted to include thermal transport, to calculate the growth rate in a wide range of situations which would be stable due to the Glasser effect. In addition the same version of MARS-F can be used to calculate suitable basis-functions for calculating $\Delta'$, since the transport effects allow penetration of the required response function at the resonant surface and evaluation of the large solution there. The application of these two methods has been demonstrated for a toroidal equilibrium with favourable average curvature, obtaining close agreement between the two methods. Thus these methods provide an alternative method, or device, for determining $\Delta'$ in toroidal geometry, complementing other approaches [7, 8]. Of course, the dispersion relation (50) also describes the effect of physically relevant values of anisotropic thermal diffusivities on tearing mode stability.

**Acknowledgements**

This work was part-funded by the RCUK Energy Programme [grant number EP/I501045]. It has been carried out within the framework of the EUROfusion Consortium and has received funding from the Euratom research and training programme 2014-2018 under grant agreement No 633053.The views and opinions expressed herein do not necessarily reflect those of the European Commission. To obtain further information on the data and models underlying this paper please contact PublicationsManager@ccfe.ac.uk.

## Appendix A: Aspects of the derivation of the tearing mode equations

In this appendix we indicate the modifications arising from the addition of thermal conduction in the equation of state to the procedure carefully displayed in Appendix B of Ref. 12 that we used to obtain the resonant layer equations (12) - (14). We closely follow the notation of Ref. 12 and frequently refer to the equation numbers of that appendix.

The key difference is to recognise that $\nabla.\xi$ must be allowed to have a contribution in $0(1)$ for consistency with the thermal conduction model. The requirement from the induction equation that $\nabla.(B^2\xi_\perp) \sim \varepsilon^2$ implies

$$(\nabla.\xi) = \mathbf{B}.\nabla\left(\nu^{(0)}/B\right) - \mathbf{B}\times\nabla V.\nabla\left(1/B^2\right)\mu^{(0)} + \varepsilon^2(\nabla.\xi)^{(2)}, \text{ where } \mathbf{B}\times\nabla V.\nabla\left(1/B^2\right) = \mathbf{B}.\nabla\sigma/P' \tag{A.1}$$

Consequently eqns (B.4), (B.8) and (B.9) of Ref.12, which result in the form (B.21) for $\nu^{(0)}$, are inappropriate. Rather $\nu^{(0)}$ is determined through the equation of state (6) and the parallel component of the momentum equation (4). In leading order of eqn. (6), we learn

$\partial\delta p^{(4)}/\partial\zeta = 0$, while in $0(\varepsilon^2)$

$$q\gamma\psi' P\frac{\partial}{\partial\zeta}\left(\frac{\nu^{(0)}}{B^2} - \frac{\sigma\mu^{(0)}}{P'}\right) = \chi_\parallel\psi'\frac{\partial}{\partial\zeta}\left(\frac{1}{B^2}\frac{\partial\delta p^{(4)}}{\partial\zeta} + \frac{\Lambda x}{\phi B^2}\frac{\partial\delta p^{(2)}}{\partial u} + P'\frac{\partial}{\partial\zeta}\left(\frac{b^{(4)}}{B^2}\right)\right) \tag{A.2}$$

Imposing the periodicity condition in $\zeta$ provides an expression for $\partial\delta p^{(4)}/\partial\zeta$. Finally, in $0(\varepsilon^4)$ we obtain, after annihilating $\delta p^{(6)}$ by averaging over $\zeta$,

$$q\left[\delta p^{(2)} + P'\xi^{(2)} + \gamma P\left\langle(\nabla.\xi)^{(2)}\right\rangle\right] = \chi_\perp\left\langle|V|^2\right\rangle\frac{d^2\delta p^{(2)}}{dx^2} + \chi_\parallel\frac{\Lambda^2x^2}{\phi^2}\frac{\partial}{\partial u}\left(\frac{1}{B^2}\frac{\partial\delta p^{(2)}}{\partial u}\right)$$
$$+\chi_\parallel\frac{\Lambda x}{\phi}\frac{\partial}{\partial u}\left\langle\frac{b^{(4)}}{B^2}\right\rangle + \chi_\parallel\frac{\Lambda x}{\phi}\frac{\partial}{\partial u}\left\langle\frac{1}{B^2}\frac{\partial\delta p^{(4)}}{\partial\zeta}\right\rangle$$

(A.3)

Inserting the result (A.3) for $\partial \delta p^{(4)}/\partial \zeta$ provides an expression for $\left\langle (\nabla.\xi)^{(2)} \right\rangle$ , involving $v^{(0)}$, to substitute in eqn. (B.11) of Ref.12. To determine $v^{(0)}$ we note that the parallel momentum equation, eqn. (B.11) of Ref. 12, provides a second expression for $\partial \delta p^{(4)}/\partial \zeta$; it is by equating these that we obtain a form for $v^{(0)}$, differing from eqn. (B.21) of Ref. 12 and exhibiting a competition between sound waves and parallel thermal conduction as a result of our new ordering.

With these modified expressions for $\partial \delta p^{(4)}/\partial \zeta$ and $v^{(0)}$ the procedure to obtain eqns. (12) – (14) follows that of Ref. 12.

## Appendix B: Dispersion relation for H $\cong$ 0 and arbitrary $D_R$

We consider the case $H \sim \delta$ when the constant $\Psi$ approximation holds. In this treatment we are no longer restricted to small values for $D_R$. Thus we have

$$\Psi^{(0)}_{XX} = 0 \tag{B.1}$$

so that

$$\Psi^{(0)} = \Psi_0 + \Psi_1 X, \tag{B.2}$$

and

$$\Psi^{(1)}_{XX} = Q\left(\Psi^{(0)} - X\Xi^{(0)}\right) + HT_X \tag{B.3}$$

We can then calculate $\Delta$:

$$\Delta = \frac{V_s}{X_0 \Psi^{(0)}} \left[\Psi^{(1)}_X\right]_{-\infty}^{+\infty} = \frac{V_s}{X_0 \Psi^{(0)}} \left( \int_{-\infty}^{+\infty} dX\left(\Psi^{(0)} - X\Xi^{(0)}\right) + HT^{(0)} \Big|_{-\infty}^{+\infty} \right) \tag{B.4}$$

The term in $T^{(0)}$ evaluated at $\pm\infty$ vanishes, so H plays no role. Using eqn. (13), this can be written

$$\Delta = -\frac{V_s}{X_0 \Psi^{(0)}} \int_{-\infty}^{+\infty} \frac{dX}{X} \left( Q^2 \Xi^{(0)}_{XX} + D_R T^{(0)} \right) \tag{B.5}$$

We first follow the approach in Ref. 5. Thus, in terms of the variable y, $T^{(0)}$ satisfies eqn. (19). This inhomogeneous equation can be solved in terms of the parabolic cylinder functions [20] and a solution odd in y, vanishing at $y = +\infty$, has the form:

$$T^{(0)}(y) = -\sqrt{\pi}\, \frac{\Gamma(3/4)}{\Gamma(1/4)} \left( \frac{\hat{\chi}_{\parallel}}{\hat{\chi}_{\perp}} \right)^{1/4} y \Psi^{(0)} \tag{B.6}$$

at small y. The equation for $\Xi^{(0)}$ becomes

$$\Xi^{(0)}_{zz} - z^2\Xi^{(0)} = -\frac{1}{Q^{5/4}}\left[Q + \sqrt{\pi}\frac{\Gamma(3/4)}{\Gamma(1/4)}\left(\frac{\hat{\chi}_{\parallel}}{\hat{\chi}_{\perp}}\right)^{1/2} D_R\right]\Psi^{(0)}z \tag{B.7}$$

Thus we can write

$$\Xi^{(0)} = -\frac{1}{Q^{5/4}}\left[Q + \sqrt{\pi}\frac{\Gamma(3/4)}{\Gamma(1/4)}\left(\frac{\hat{\chi}_{\parallel}}{\hat{\chi}_{\perp}}\right)^{1/2} D_R\right]\Psi^{(0)}F(z) \tag{B.8}$$

and

$$T^{(0)}(y) = -\left(\frac{\hat{\chi}_{\parallel}}{\hat{\chi}_{\perp}}\right)^{1/4} F(y) \tag{B.9}$$

where

$$F(x)_{xx} - x^2F(x) = -x \tag{B.10}$$

Using the results [5]

$$\int_{-\infty}^{\infty}\frac{dz}{z}F_{ZZ}(z) = -2\pi\frac{\Gamma(3/4)}{\Gamma(1/4)}, \qquad \int_{-\infty}^{\infty}\frac{dy}{y}F(y) = \frac{\pi^{3/2}}{2}, \tag{B.11}$$

we finally have

$$\Delta(Q) = 2\pi Q^{5/4}\left(\frac{V_s}{X_0}\right)\frac{\Gamma(3/4)}{\Gamma(1/4)}\left[1 + \sqrt{\pi}\left(\frac{\hat{\chi}_{\parallel}}{\hat{\chi}_{\perp}}\right)^{1/2}\frac{D_R}{Q}\frac{\Gamma(3/4)}{\Gamma(1/4)}\right] - \frac{\pi^{3/2}}{2}\left(\frac{V_s}{X_0}\right)\left(\frac{\hat{\chi}_{\parallel}}{\hat{\chi}_{\perp}}\right)^{1/4} D_R \tag{B.12}$$

This result is a mild generalisation of that in Ref. 5 and recovers eqn. (51).

An alternative approach involving Fourier transforms allows an exact solution to the problem and provides some light on the more complicated mathematics for the case with $H \neq 0$ considered in section 3, which also uses Fourier transform techniques, but there on third order differential equations, rather than the present second order ones.

We first express eqn. (B.4) in terms of the variable $z$

$$\Delta = \frac{V_s}{X_0} Q^{5/4} \int_{-\infty}^{\infty} dz \left(1 - Q^{1/4} z \Xi^{(0)}\right) \qquad \text{(B.13)}$$

and introduce the Fourier transform $\hat{\Xi}(k) = \int_{-\infty}^{\infty} dz\, e^{ikz} \Xi(x)$ so that eqn. (B.13) becomes

$$\Delta = -\frac{V_s}{X_0} Q^{5/4} \left( \delta(k) + iQ^{1/4} \left.\frac{d\hat{\Xi}^{(0)}}{dk}\right|_0 \right) \qquad \text{(B.14)}$$

The equation for $\hat{\Xi}^{(0)}$ is

$$\hat{\Xi}^{(0)}_{kk} - k^2 \hat{\Xi}^{(0)} = \frac{2\pi i}{Q^{1/4}} \frac{d\delta(k)}{dk} - \frac{D_R}{Q^{3/2}} \hat{T}(k) \qquad \text{(B.15)}$$

However the equation for $\hat{T}(k)$ is

$$\hat{T}^{(0)}_{\kappa\kappa} - \kappa^2 \hat{T}^{(0)} = \frac{2\pi i \lambda}{Q^{1/4}} \frac{d\,\delta(\kappa)}{d\kappa} \qquad \text{(B.16)}$$

where $k = \lambda\kappa$. Equation (B.16) implies the boundary condition $\hat{T}(0) = \pi i \lambda / Q^{1/4}$ so that the solution of eqn. (B.16) vanishing at large $\kappa$ is

$$\hat{T}^{(0)}(\kappa) = \frac{2^{3/4} i \lambda \Gamma(3/4)}{Q^{1/4}} s^{1/4} K_{1/4}(s) \qquad \text{(B.17)}$$

where $K_{1/4}(s)$ is a modified Bessel function of the second kind and $s = \kappa^2/2$. Likewise eqn. (B.15) yields $\hat{\Xi}^{(0)}(0) = \pi i / Q^{1/4}$ and a solution

$$\hat{\Xi}^{(0)} = \left( \frac{2^{3/4} i \Gamma(3/4)}{Q^{1/4}} t^{1/4} K_{1/4}(t) + t^{1/4} K_{1/4}(t) \int_0^t dt\, I_{1/4}(t) R(t) + + t^{1/4} I_{1/4}(t) \int_t^{\infty} dt K_{1/4}(t) R(t) \right) \mathrm{sgn}(k)$$

(B.18)

where $I_{1/4}(t)$ is a modified Bessel function of the first kind and

$$R(t) = i \frac{D_R \Gamma(3/4)}{2^{3/4} Q^{7/4} \lambda^{1/2}} K_{1/4}\left( \frac{t}{\lambda^2} \right) \qquad \text{(B.19)}$$

We can now calculate $d\hat{\Xi}^{(0)}/dk$ at $k = 0$:

$$\left.\frac{d\hat{\Xi}^{(0)}}{dk}\right|_0 = \frac{\pi i}{Q^{1/4}} \delta(k) - 2\pi i \frac{\Gamma(3/4)}{\Gamma(1/4) Q^{1/4}} \left( 1 + \sqrt{\pi} \frac{D_R}{Q^{3/2}} \lambda^2 \frac{\Gamma(3/4)}{\Gamma(1/4)} \right) + i \frac{\pi^{3/2}}{2^{1/4}} \frac{D_R}{Q^{7/4}} \lambda \qquad \text{(B.20)}$$

Substituting this result into eqn. (B.14), we recover eqns. (51) and (B.12). In deriving eqn. (51), the extraction of the large z behaviour of $\hat{\Xi}^{(0)}$ involved the region $k << 1$; we note that eqn. (B.20) also involves the same region of k.

One can readily extend the analysis to arbitrary values of $\lambda$. The terms proportional to $D_R$ in $\Delta'$ arise from the last integral in eqn. (B.18) which can be evaluated for all values of $\lambda$. They are proportional to the function $G(\lambda)$:

$$G(\lambda) = \lambda^{-1/2} \int_0^\infty dt K_{1/4}(t) K_{1/4}(t/\lambda^2), \qquad \text{(B.21)}$$

where

$$\begin{aligned} G(\lambda) &= \frac{\pi\lambda^2}{4} \Gamma(1/4)\Gamma(3/4)\, {}_2F_1(3/4, 1/2, 1; 1-\lambda^4); \quad \lambda < 1 \\ G(\lambda) &= \frac{\pi}{4\lambda} \Gamma(1/4)\Gamma(3/4)\, {}_2F_1(3/4, 1/2, 1; 1-\lambda^{-4}); \quad \lambda > 1, \end{aligned} \qquad \text{(B.22)}$$

which is continuous at $\lambda = 1$. The function $G(\lambda)$ increases monotonically with $\lambda$ and asymptotes to the constant value $G(\lambda) = \Gamma(5/4)/\sqrt{\pi}\Gamma(3/4)$. This means that in the limit $\lambda >> 1$,

$$\Delta(Q) = 2\pi \frac{V_s}{X_0} Q^{5/4} \frac{\Gamma(3/4)}{\Gamma(1/4)} \left[ 1 - \frac{1}{2^{5/2}\pi^{3/2}} \frac{\Gamma(1/4)}{\Gamma(3/4)} \frac{D_R}{Q^{3/2}} \right]. \qquad \text{(B.23)}$$

## Figure Captions

**Fig. 1:** The growth rate $\gamma$ normalised to the Alfven time $\tau_A$ as a function of Lundquist Number, S, as calculated by MARS-F for the resistive MHD model (dashed line with squares) and thermal conduction model (solid line with circles), showing how the resistive case is stabilised at large S. The resistive MHD model assumes a uniform profile for the plasma resistivity. The thermal conduction model assumes $\chi_\parallel/\chi_\perp = 10^2$ in this scan. The tearing mode becomes stable at $S = 10^9$ for this plasma due to the Glasser effect.

**Fig. 2**: $\Delta'$ as a function of $\chi_\parallel/\chi_\perp$ inferred from MARS for the thermal transport model calculated using the general dispersion relation for H > 0 (solid line with squares) and the one for H = 0 (dashed lines with circles): the small value of H for the particular equilibrium makes them lie very close to each other. The Lundquist number is $S = 10^9$ in this scan.

**Fig. 3**: $\Delta'$ as a function of S as inferred from the thermal transport model for $\chi_\parallel/\chi_\perp = 10^2$ (solid line with circles) and using the dispersion relation of the resistive MHD model from Ref. 2 (dashed line with squares); the two methods recover the same value for $\Delta'$ at large values of S.

**Fig. 4**: Perturbed radial magnetic field response function (labelled by poloidal harmonic mode number) as functions of a 'radial' flux surface co-ordinate, s, calculated by MARS-F for the resistive MHD model, showing the screening of the m = 2 component at the resonant surface. The Lundquist number is $S = 10^9$ in this example.

**Fig. 5**: Perturbed radial magnetic field response function (labelled by poloidal harmonic mode number) as functions of the 'radial' flux surface co-ordinate, s, calculated by MARS-F for the thermal transport model, showing the absence of screening of the m = 2 component at the resonant surface. The Lundquist number is $S = 10^9$ in this example.

**Fig. 6**: Comparison of values of $\Delta'$ as a function of $\chi_{\parallel}/\chi_{\perp}$ calculated using: (i) the resistive MHD dispersion relation of Ref. 2 (dashed line), (ii) the thermal transport model (line with circles) and (iii) the basis-function method (dash-dot line with diamonds). The Lundquist number is $S = 10^9$ in this example.

**Fig. 7**: Comparison of values of $\Delta'$ as a function of S: calculated using (i) the resistive MHD dispersion relation of Ref. 2 (dashed line with squares), (ii) the thermal transport model (line with circles) and (iii) the basis-function method (dash-dot line with diamonds); the resistive MHD model is unstable for this range of S.

Fig 1

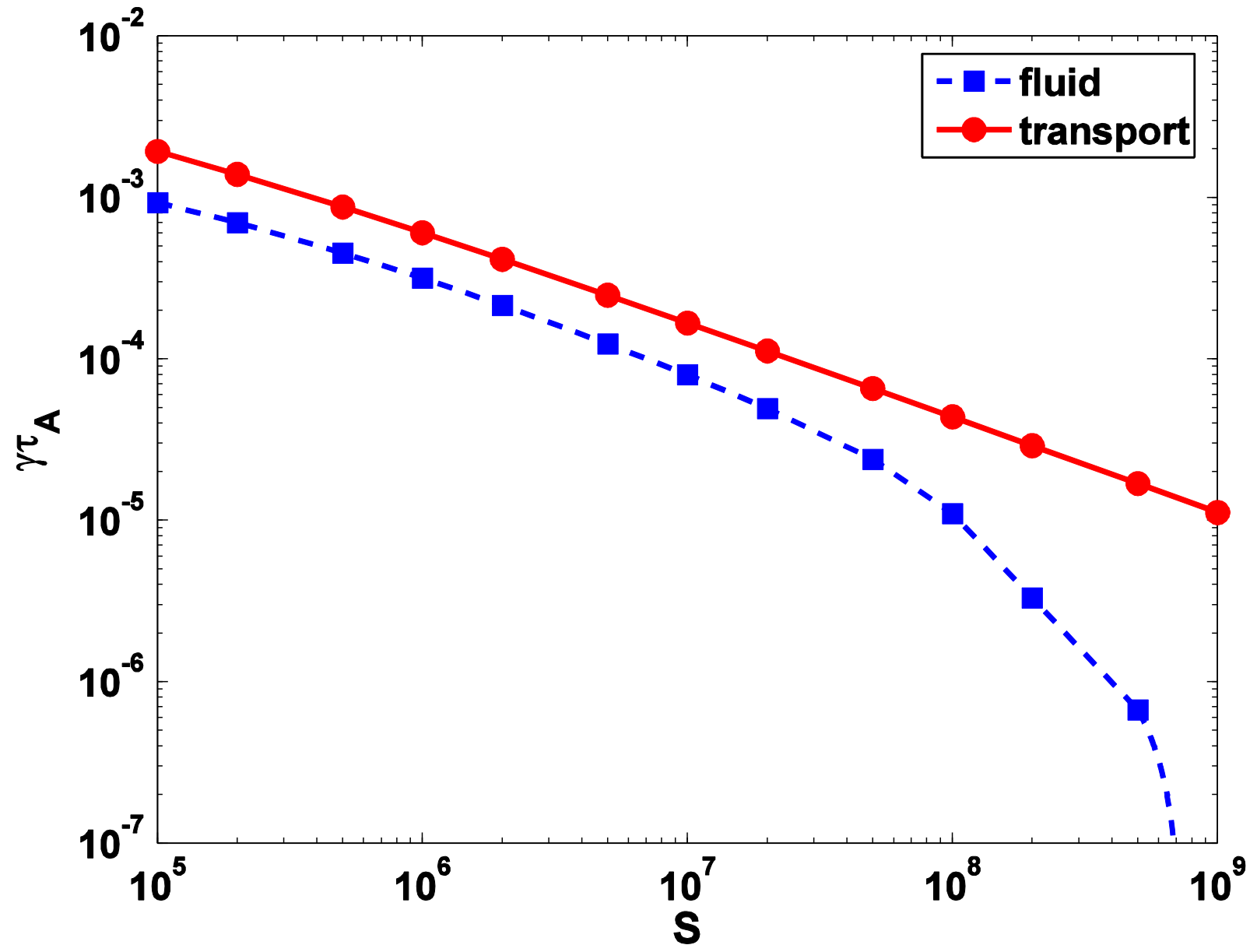

Fig 2

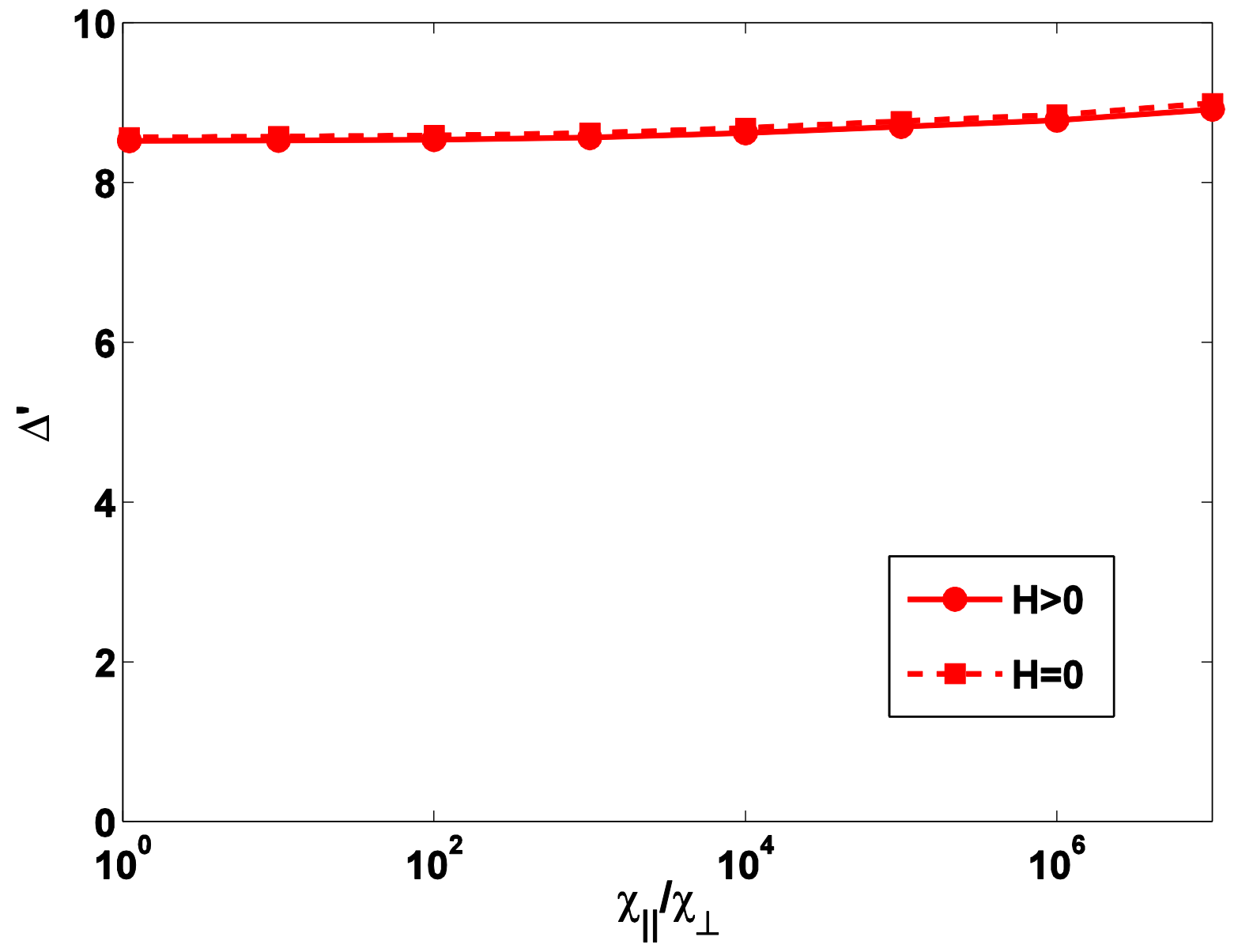

Fig 3

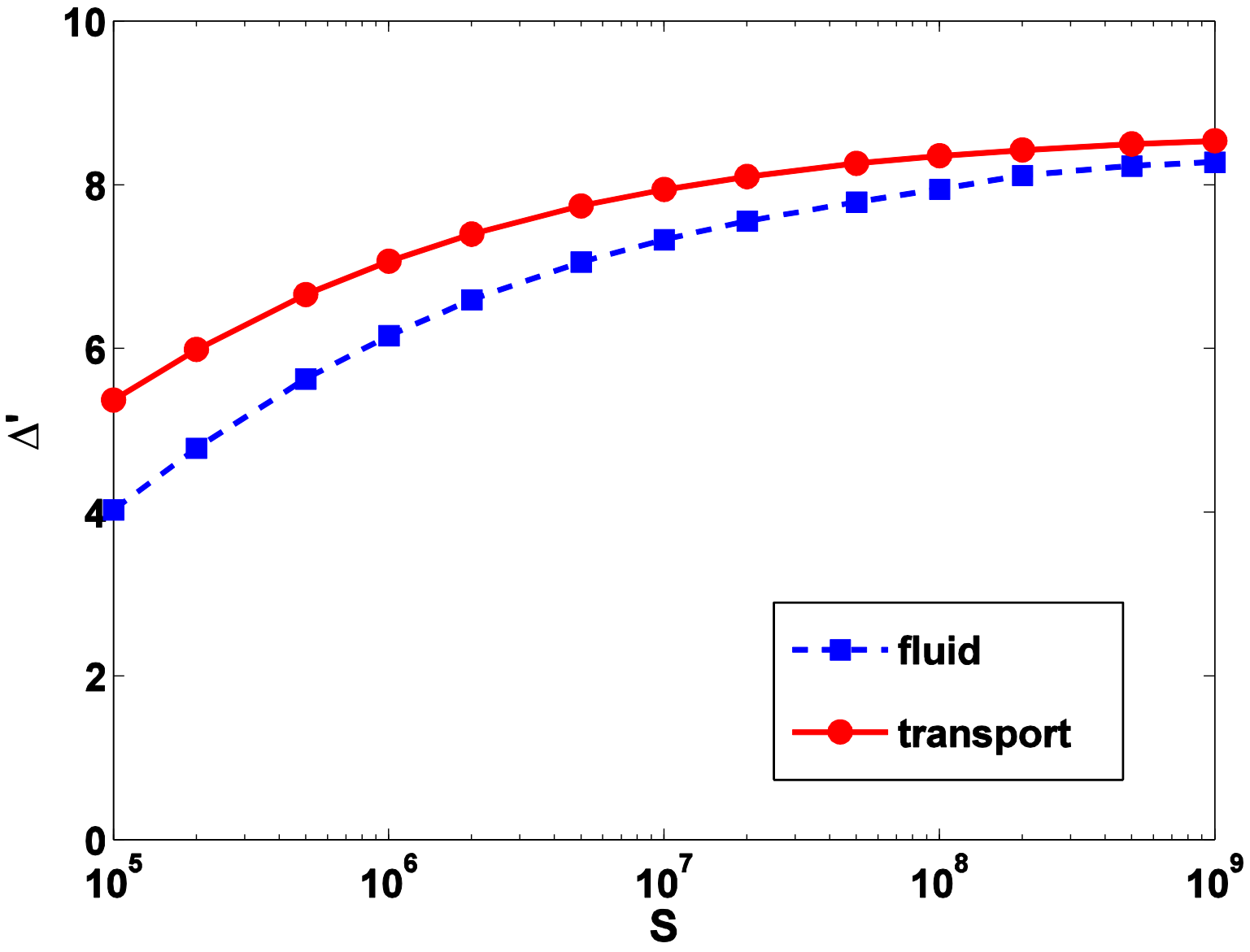

Fig 4

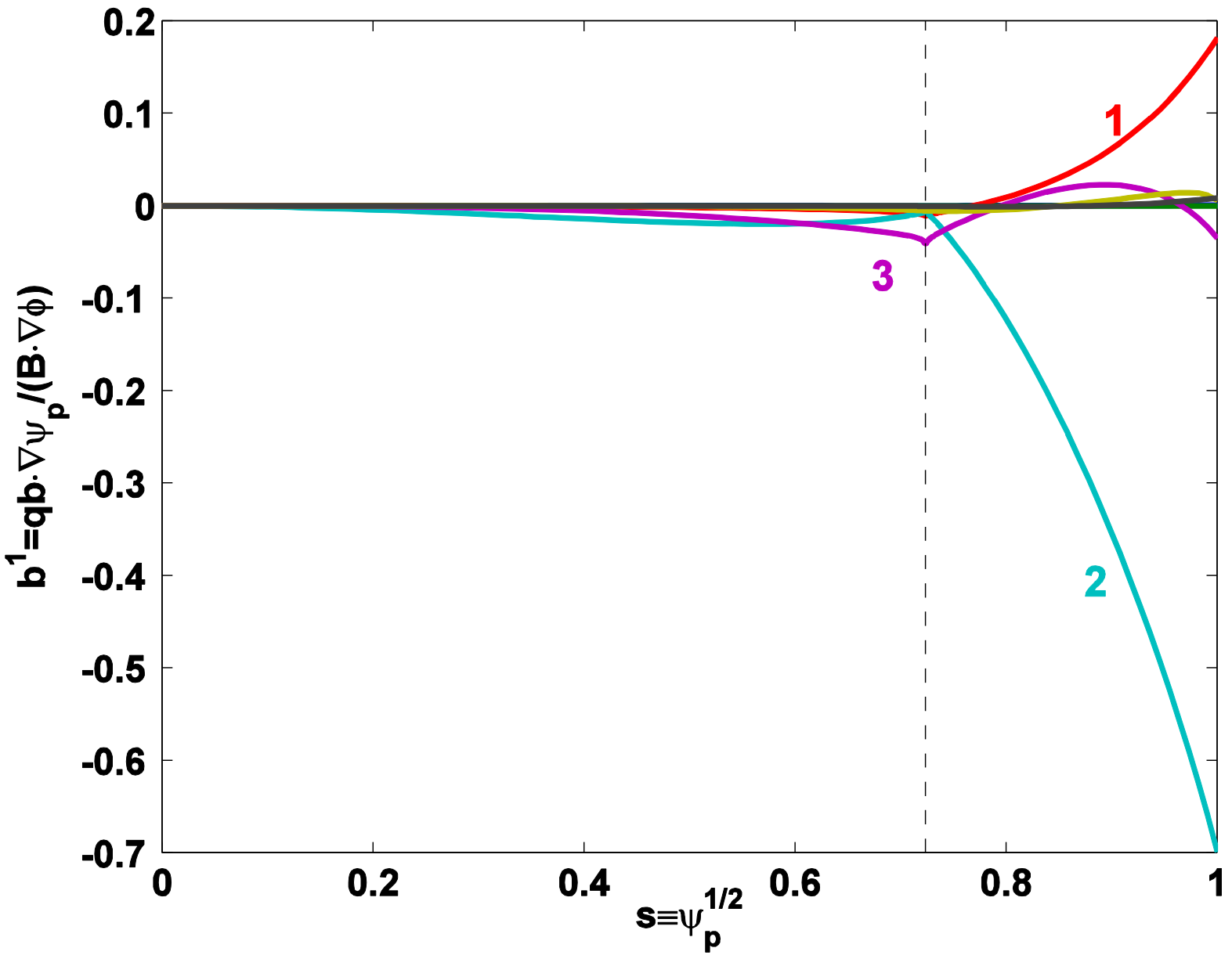

Fig 5

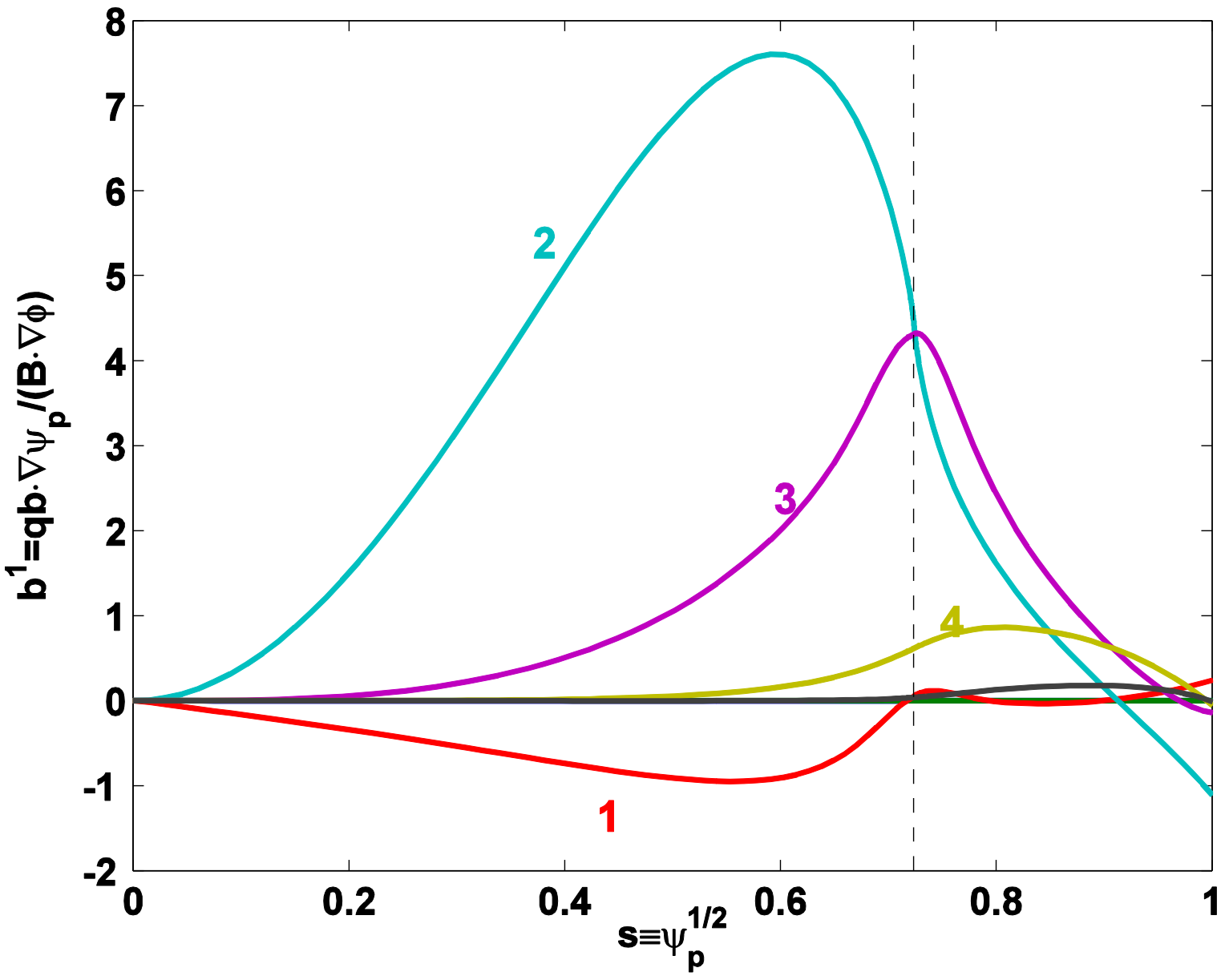

Fig 6

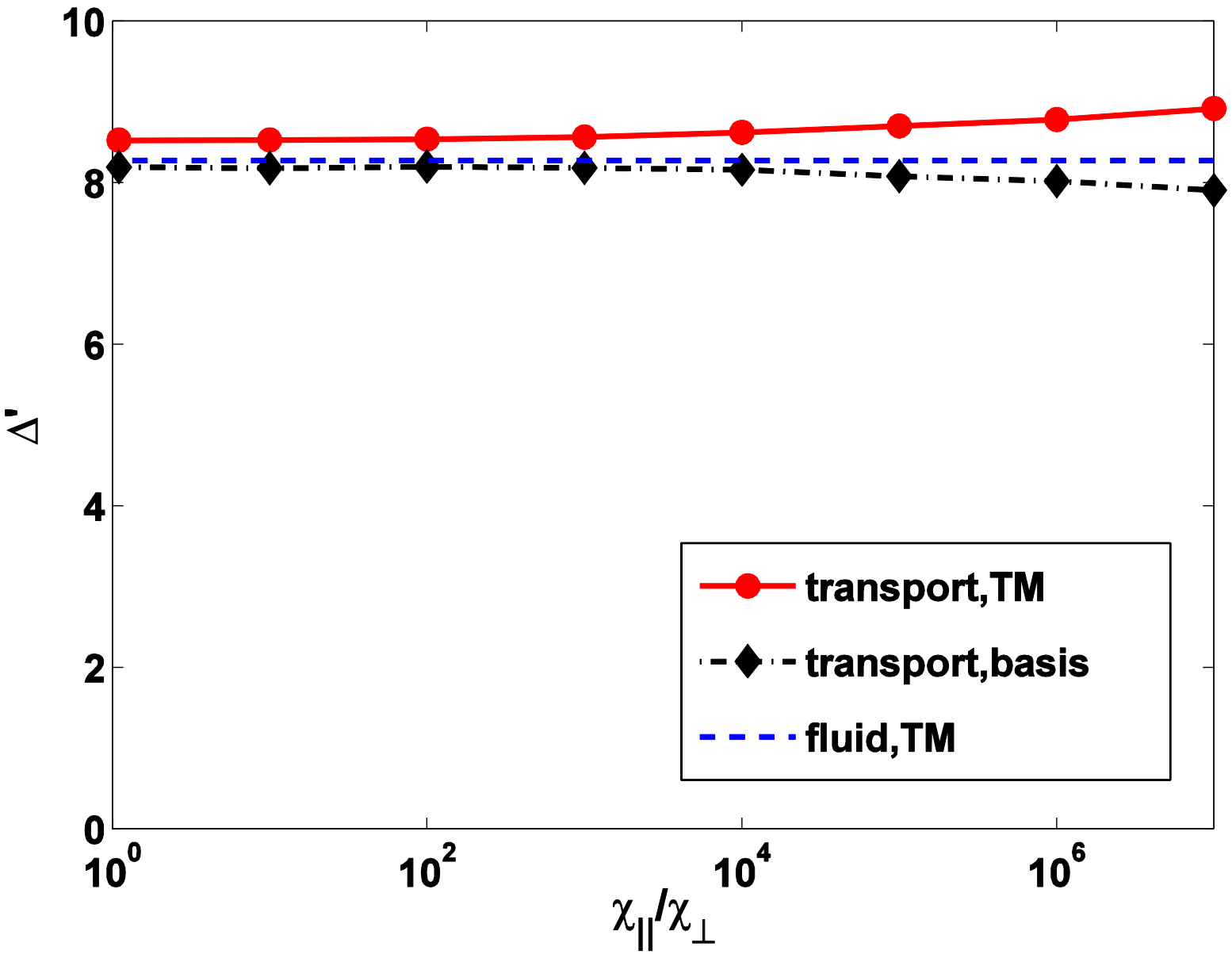

Fig 7

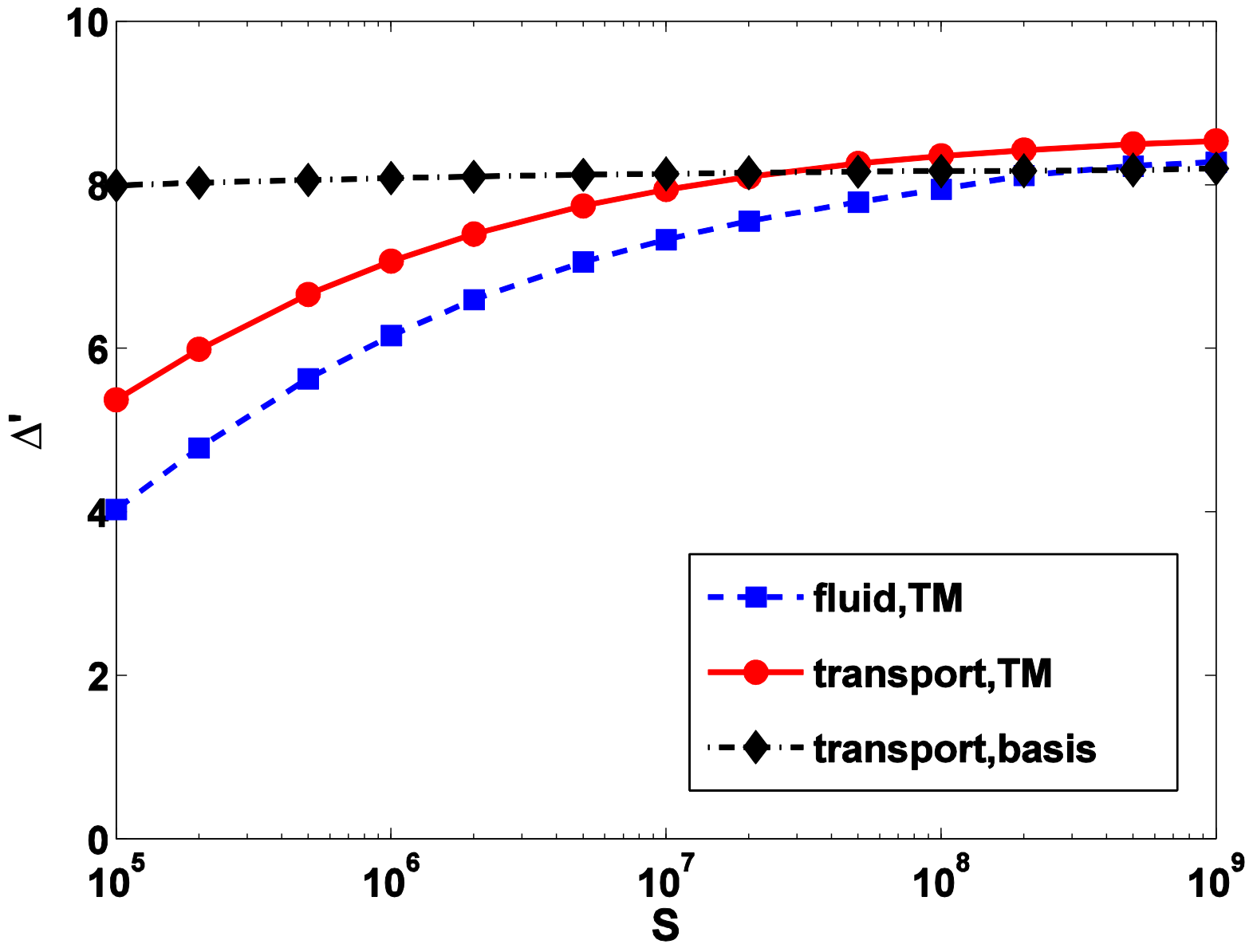